\documentclass[acmsmall,nonacm]{acmart}

\usepackage{graphicx} 

\begin{document}

\title[Strategies for engaging clinical participants in the co-design of software for healthcare domains]{Strategies for engaging clinical participants in the co-design of software for healthcare domains}

% Getting rid of unncessary frontmatter and formatting
\setcopyright{none}
\settopmatter{printacmref=false, printccs=false, printfolios=false}
\authorsaddresses{}
\acmVolume{}
\acmNumber{}
\acmArticle{}
\acmYear{}
\acmMonth{}
\acmPrice{}
\acmDOI{}
\acmISBN{}

\author{Marceli Wac}
\email{m.wac@bristol.ac.uk}
\orcid{0000-0002-1986-0401}

\author{Raul Santos-Rodriguez}
\email{enrsr@bristol.ac.uk}
\orcid{0000-0001-9576-3905}

\author{Chris McWilliams}
\email{chris.mcwilliams@bristol.ac.uk}
\orcid{0000-0003-3816-5217}

\affiliation{
  \institution{University of Bristol}
  \city{Bristol}
  \country{United Kingdom}
}

\author{Christopher Bourdeaux}
\email{christopher.bourdeaux@uhbw.nhs.uk}
\orcid{0000-0001-6620-6536}

\affiliation{
  \institution{University Hospitals Bristol and Weston National Health Service Foundation Trust}
  \city{Bristol}
  \country{United Kingdom}
}

\renewcommand{\shortauthors}{Wac, et al.}

\begin{abstract}

\textit{
Co-design is an effective method for designing software, but implementing it within the clinical setting comes with a set of unique challenges.
This makes recruitment and engagement of participants difficult, which has been demonstrated in our study.
Our work focused on designing and evaluating a data annotation tool, however, different types of interventions had to be carried out due to poor engagement with the study.
We evaluated the effectiveness and feasibility of each of these strategies, their applicability to different stages of co-design research and discussed the barriers to participation present among participants from a clinical background.
}

\end{abstract}


% Commented out for the purpose of this submission
% \keywords{software engineering, co-design, clinical participant, recruitment, engagement}

\begin{teaserfigure}
  \includegraphics[width=\textwidth]{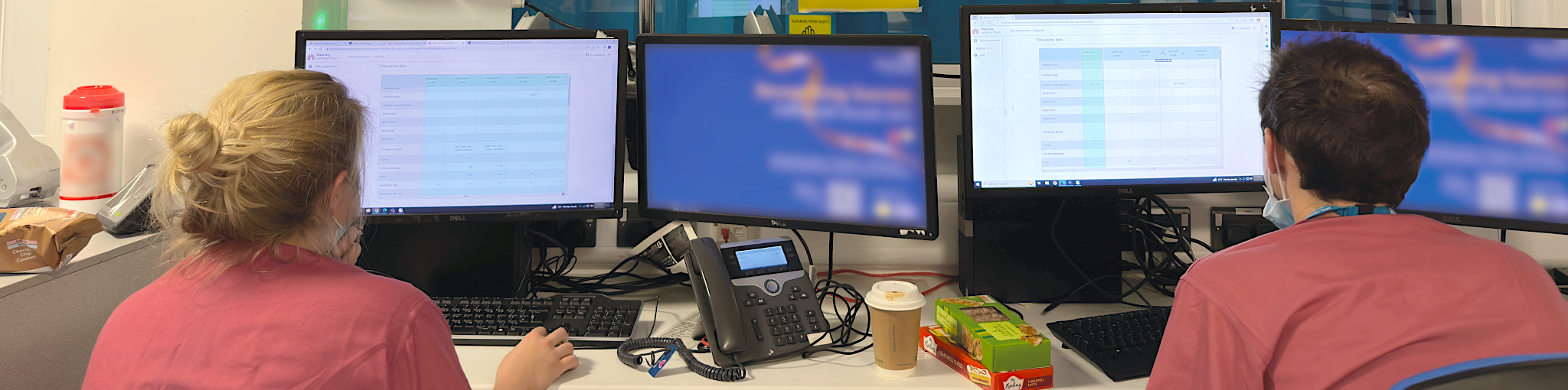}
  \caption{Study participants annotating data using our tool during a workshop in the intensive care unit.}
  \Description{Clinical staff annotating data using the bespoke tool. Participants had the opportunity to label time-series data during the workshop held in the intensive care unit.}
  \label{fig:teaser}
\end{teaserfigure}

\maketitle

\section{Introduction}

The constant advances in the field of informatics, and in particular data science and artificial intelligence, have brought on a period of rapid growth in research applying those technologies to healthcare settings \cite{Noorbakhsh-Sabet2019,Davenport2019}.
Such research frequently involves stakeholders from the clinical domain, such as doctors, nurses and consultants in a variety of ways through a participatory design or a co-design process.
In human-computer interaction (HCI), these can include user workshops exploring the ideas around a certain problem, formal interviews to elicit requirements, as well as design and evaluation of user interfaces for the proposed solution.
Incorporating end-users in the design process within healthcare has been shown to result in improved effectiveness of the final product, but the unique characteristics of this environment make conducting co-design studies more challenging \cite{Bowen2013}.
Clinicians are oftentimes experts in a specific area of healthcare; they work with strict schedules in stressful environments and make decisions that affect lives of others on a daily basis.
This suggests, that there are barriers to the recruitment of participants and challenges related to retaining their interest, which are inherent to the healthcare settings.
The often lengthy nature of software development, need for regulatory approvals, privacy and security requirements and infrastructural contrstraints in the clinical domain further complicate this process.
This combination creates challenges which make it critical to ensure that the participants invited to take part in the research will not only engage with the study, but remain involved throughout its course.
Our work presents the current progress in a study focusing on the creation and evaluation of a bespoke data annotation tool which has been co-designed with clinicians, the challenges associated with the recruitment of the participants, their engagement throughout the study as well as strategies implemented to improve participation.

\section{Methodology}

The study took place at the University Hospitals Bristol and Weston NHS Trust (UHBW) in the UK in the first quarter of 2023 and focused on co-designing and implementing a data annotation platform purpose-built for labelling clinical time-series data.
The design process involved two stages: an early, exploratory analysis of how clinicians approach data annotation (S1), and test and evaluation of the digital prototype developed based on the feedback from the previous stage (S2).
The recruitment for both S1 and S2 was set to take the form of an invitational e-mail sent out to the eligible members of the trust.
All participants involved in both S1 and S2 held positions at the intensive care unit (ICU) at the time of the study.

The first stage (S1) focused on establishing the most efficient way for capturing the annotations, the different ways the annotations can be marked and the overall approach to the annotation problem.
In particular, S1 was aimed at informally extracting the functional requirements for the software tool.
To that extent, participants were presented with a print-out of a table containing the respiratory data and patient vitals, and tasked with annotating periods of time during which weaning from mechanical ventilation takes place.
S1 took place in a form of a single, in-person workshop held at UHBW.
Based on the results from S1, a software platform encapsulating the data annotation workflow was developed and deployed on Amazon Web Services (AWS).
The platform exposed a web interface which allowed participants to enrol in the study, register for an account and perform the annotation task on a subset of data assigned to them during registration.
The data presented to participants came from the Medical Information Mart for Intensive Care dataset (MIMIC-IV v2.1) \cite{Alistair2022}.
The second stage focused on testing the tool and required participants to annotate patient admissions.
This involved inspecting the time-series data presented in a table which has been purposefully styled to resemble the interface of the clinical information system present at UHBW, selecting start and end points on the timeline to mark a region of interest and specifying label details (parameters suggesting label presence, explanation of their significance and confidence in label accuracy).
Each admission could be annotated with multiple labels and upon submitting the labels, the platform would automatically redirect the participant to the next assigned admission.
To maximise the accessibility, the platform was kept available for 28 days, allowing participants to take part asynchronously and remotely.
Additionally, a detailed help page describing the interface and annotation flow was embedded on the platform and key instructions were extracted and placed at the top of the labelling interface to help alleviate usage-related barriers.

\section{Results}

\begin{figure}%[h!]
  \includegraphics[width=\textwidth]{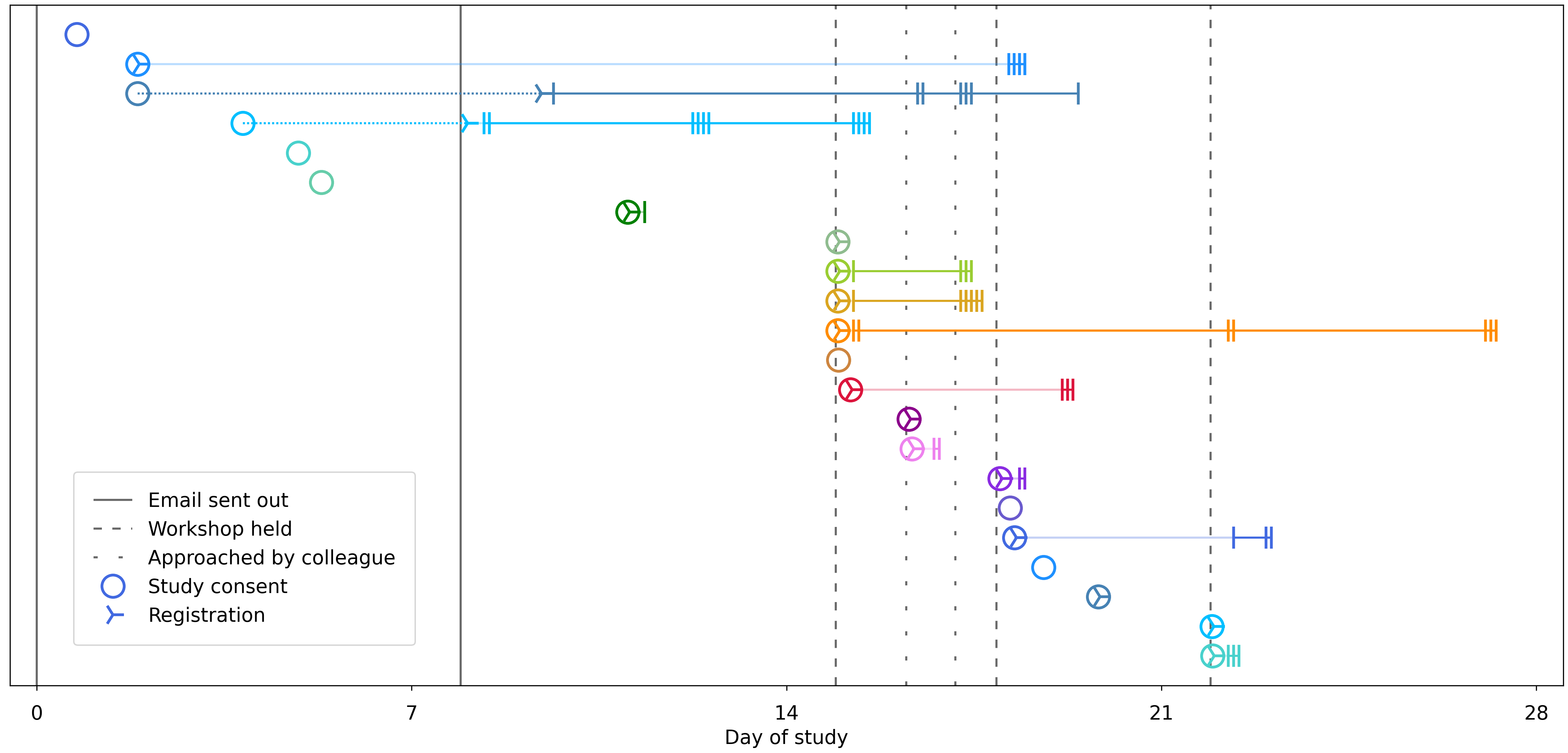}
  \caption{Participant engagement throughout the study has improved following our interventions. Workshops were most effective at recruiting new participants while approach by colleague yielded most annotations.}
  \Description{A timeline of the second stage of the study showcasing participant engagement following each intervention. The engagement has improved following each intervention. Workshops were most effective at recruiting new participants while approach by colleague yielded most annotations.}
  \label{fig:responses}
\end{figure}

\begin{figure}%[h!]
    \begin{minipage}{0.49\textwidth}
        \includegraphics[width=\textwidth]{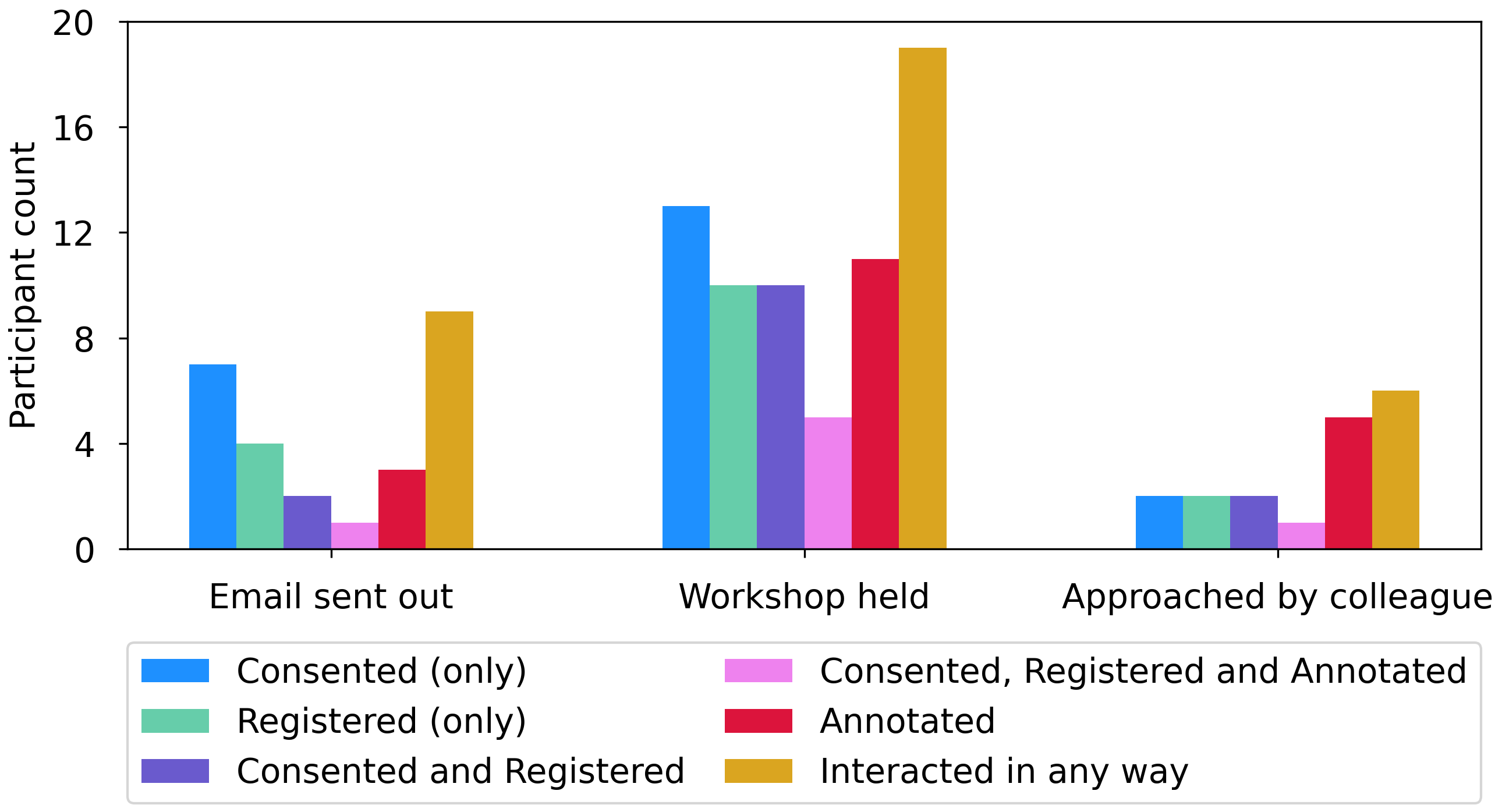}
        \caption{Participants engagement following each intervention. Workshops generated higest numbers of consent signatures, registrations and annotating participants; approach by colleague was primarily effective at engaging participants with annotation.}
        \label{fig:responses_category}
    \end{minipage}
    \hfill
    \begin{minipage}{0.49\textwidth}
        \includegraphics[width=\textwidth]{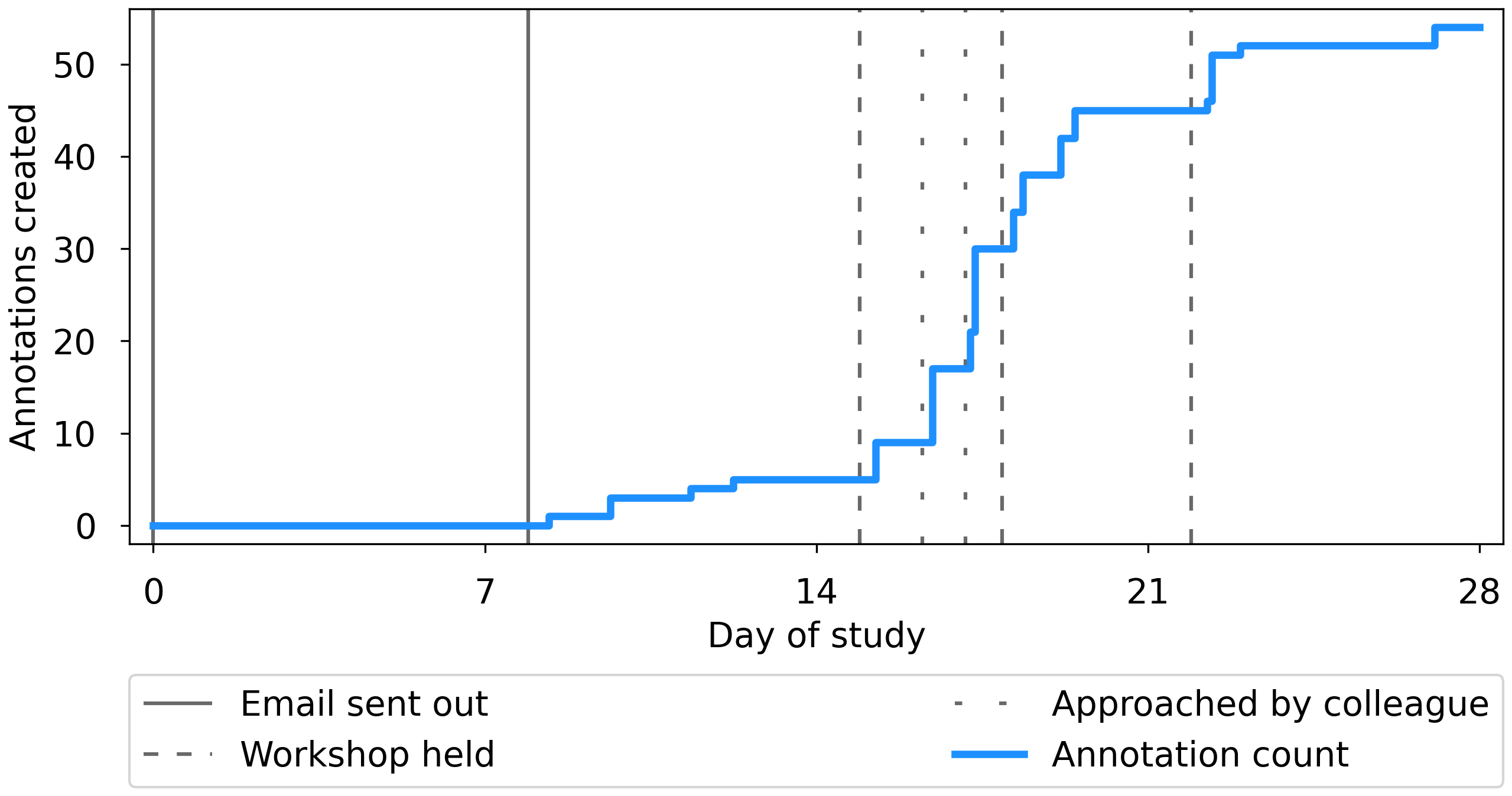}
        \caption{Cumulative annotation count throughout the study. Initial lack of annotations has improved following email intervention and workshops, but approach by the colleagues resulted in the largest growth of annotations created using our tool.}
        \label{fig:responses_annotations}
    \end{minipage}
\end{figure}

In total, 7 participants took part in S1 during the single workshop and 22 participants took part in S2 over 28 days (22 participants consented, 16 registered, 12 created annotations).
The initial email inviting participants to take part in the study was effective for S1, where the targetted participant count was smaller.
Applying the same strategy to S2 was not as effective and produced disappointing results; it generated 6 consent signatures, 1 registration and no annotations (see Fig. \ref{fig:responses_category}).
These unsatisfactory results necessitated intervention which would recruit additional participants.
Re-sending the email and prompting prospective participants to sign up again resulted in further 1 consent signature, 3 registrations and 8 annotations.
While this improved the participant count, it had minimal effect on their engagement in the annotation task, indicating the need for further interventions targetting participant engagement specifically.
The limited response to the call for participation via email suggested that a more personal approach could be more feasible.
Following the end of the second week of the study, three 2-hour workshops discussing data science within healthcare and outlining the study were held in the ICU at UHBW.
This contextualised the study to participants and resulted in additional 13 consent signatures, 10 registrations and 31 annotations.
Workshops led to more annotations created by existing participants, but those who enrolled immediately afterwards presented limited engagement.
To encourage more continuous participation, we decided to focus on utilising existing community at the trust and onboard a clinician who would signpost and promote the study among their peers.
Approach by the colleague generated further 2 consent signatures, 2 registrations and 17 annotations.
The timeline of interventions and participant engagement in the second stage of the study is depicted on the Fig. \ref{fig:responses}.

Overall, the most effective method of engaging with participants (generating consent form signatures, registrations and participation in the annotation task) have been the workshops held in-person at the ICU (see Fig. \ref{fig:responses_category}).
Approach by colleague led to the largest number of created annotations following the intervention (see Fig. \ref{fig:responses_annotations}); despite being ineffective as a recruitment strategy, it had the highest conversion rate (percentage of participants who consented, registered and annotated data).
Recruitment via email resulted in participants registering on the system, but not engaging in the annotation task (see Fig. \ref{fig:responses_category}).

\section{Discussion}

Recruitment and engagement of participants during the second stage of the study have shown to be challenging despite the efforts made to accommodate the schedules of clinicians and making the task and tool accessible and intuitive.
The primary barrier to undertaking this research has been the number of ethical and organisational approvals that needed to be granted prior to the study taking place and the recruitment of participants from the clinical background.
Designing the study to fit with the individual schedules of participants has provided mixed results; it allowed for the task to be completed asynchronously which has been utilised by the participants but disincentivised the participation in the first place, leading to poor engagement.
Despite incentives (refreshments provided during workshops) and running the workshop as a structured learning event, which are required for the trainees' portfolio, participants enrolled but remained mostly uninterested in the task. 
This could be attributed to the high frequency at which research is being conducted at this particular trust, as well as differences in incentive preferences among the participants from a clinical background.

Different types of interventions had varying effects on the participants' involvement in the study.
Re-sending the invitational email has prompted existing participants to create an account in the tool and undertake the annotation task, but largely failed at recruiting participants.
This suggests a follow-up communication to be a viable strategy for re-engaging existing participants who have already enrolled, and a limited feasibility of repeated attempts at recruitment via email.
While the majority of the participants signed the consent forms and registered for an account following email and workshop interventions, the primary motivators for engagement with the task were both workshops and approach of the colleague.
In-person workshops focusing on both recruitment and engagement with the annotation task have proved to be effective but had short-lived effects (when the workshops ended, participants did not return to the task).
Addressing the potential participants in person allowed for the recruitment of participants who did not respond to the email, but was limited by the current availability of clinical staff.
Approach of the colleague has shown to be primarily effective in engaging existing participants, rather than recruiting new ones.
Participants have created the largest number of annotations after being prompted by a colleague, suggesting that utilising social connections to promote interest in the study can improve engagement.
We acknowledge these results might have been affected by the varying time spans between different interventions and that potentially, if given time, some participants would have engaged with the task after a longer period following the workshop (without being prompted by their colleague).

\begin{acks}
MW was funded by the EPSRC Digital Health and Care CDT EP/S023704/1.
RSR was funded by the UKRI Turing AI Fellowship EP/V024817/1.
Project received AWS Cloud Credit for Research grant.
\end{acks}

\bibliographystyle{ACM-Reference-Format}
\bibliography{library}


\begin{thebibliography}{4}

%%% the Web address.

\ifx \showCODEN    \undefined \def \showCODEN     #1{\unskip}     \fi
\ifx \showDOI      \undefined \def \showDOI       #1{#1}\fi
\ifx \showISBNx    \undefined \def \showISBNx     #1{\unskip}     \fi
\ifx \showISBNxiii \undefined \def \showISBNxiii  #1{\unskip}     \fi
\ifx \showISSN     \undefined \def \showISSN      #1{\unskip}     \fi
\ifx \showLCCN     \undefined \def \showLCCN      #1{\unskip}     \fi
\ifx \shownote     \undefined \def \shownote      #1{#1}          \fi
\ifx \showarticletitle \undefined \def \showarticletitle #1{#1}   \fi
\ifx \showURL      \undefined \def \showURL       {\relax}        \fi
% invisible to TeX
\providecommand\bibfield[2]{#2}
\providecommand\bibinfo[2]{#2}
\providecommand\natexlab[1]{#1}
\providecommand\showeprint[2][]{arXiv:#2}

\bibitem[Bowen et~al\mbox{.}(2013)]%
        {Bowen2013}
\bibfield{author}{\bibinfo{person}{Simon Bowen}, \bibinfo{person}{Kerry
  McSeveny}, \bibinfo{person}{Eleanor Lockley}, \bibinfo{person}{Daniel
  Wolstenholme}, \bibinfo{person}{Mark Cobb}, {and} \bibinfo{person}{Andy
  Dearden}.} \bibinfo{year}{2013}\natexlab{}.
\newblock \showarticletitle{How was it for you? Experiences of participatory
  design in the UK health service}.
\newblock \bibinfo{journal}{\emph{CoDesign}}  \bibinfo{volume}{9}
  (\bibinfo{date}{12} \bibinfo{year}{2013}), \bibinfo{pages}{230--246}.
\newblock
Issue 4.
\showISSN{15710882}
\urldef\tempurl%
\url{https://doi.org/10.1080/15710882.2013.846384}
\showDOI{\tempurl}


\bibitem[Davenport and Kalakota(2019)]%
        {Davenport2019}
\bibfield{author}{\bibinfo{person}{Thomas Davenport} {and}
  \bibinfo{person}{Ravi Kalakota}.} \bibinfo{year}{2019}\natexlab{}.
\newblock \showarticletitle{The potential for artificial intelligence in
  healthcare}.
\newblock \bibinfo{journal}{\emph{Future Healthcare Journal}}
  \bibinfo{volume}{6} (\bibinfo{date}{6} \bibinfo{year}{2019}),
  \bibinfo{pages}{94--98}.
\newblock
Issue 2.
\showISSN{2514-6645}
\urldef\tempurl%
\url{https://doi.org/10.7861/futurehosp.6-2-94}
\showDOI{\tempurl}


\bibitem[Johnson et~al\mbox{.}(2022)]%
        {Alistair2022}
\bibfield{author}{\bibinfo{person}{Alistair Johnson}, \bibinfo{person}{Lucas
  Bulgarelli}, \bibinfo{person}{Tom Pollard}, \bibinfo{person}{Steven Horng},
  \bibinfo{person}{Leo~Anthony Celi}, {and} \bibinfo{person}{Mark Roger}.}
  \bibinfo{year}{2022}\natexlab{}.
\newblock \bibinfo{booktitle}{\emph{MIMIC-IV (version 2.1)}}.
\newblock PhysioNet.
\newblock
\urldef\tempurl%
\url{https://doi.org/10.13026/rrgf-xw32}
\showDOI{\tempurl}


\bibitem[Noorbakhsh-Sabet et~al\mbox{.}(2019)]%
        {Noorbakhsh-Sabet2019}
\bibfield{author}{\bibinfo{person}{Nariman Noorbakhsh-Sabet},
  \bibinfo{person}{Ramin Zand}, \bibinfo{person}{Yanfei Zhang}, {and}
  \bibinfo{person}{Vida Abedi}.} \bibinfo{year}{2019}\natexlab{}.
\newblock \showarticletitle{Artificial Intelligence Transforms the Future of
  Health Care}.
\newblock \bibinfo{journal}{\emph{The American Journal of Medicine}}
  \bibinfo{volume}{132} (\bibinfo{date}{7} \bibinfo{year}{2019}),
  \bibinfo{pages}{795--801}.
\newblock
Issue 7.
\showISSN{00029343}
\urldef\tempurl%
\url{https://doi.org/10.1016/j.amjmed.2019.01.017}
\showDOI{\tempurl}


\end{thebibliography}

\end{document}